\documentclass[a4paper]{jpconf}
\usepackage{graphicx,amssymb,isotope,hyperref}
\begin{document}
\title{Internal Backgrounds in the Water Phase of SNO+}

\author{Ian Lam, for the SNO+ Collaboration}

\address{Queen's University}

\ead{ian.lam@queensu.ca}

\begin{abstract}
SNO+ is a neutrinoless double-beta decay ($0\nu\beta\beta$) search experiment using 780 tonnes of tellurium-loaded liquid scintillator. The experiment is currently collecting data in the first of three planned phases, in which the detector is filled with ultrapure water. During this phase, the cleanliness of the water can be assayed using in situ measurements of \isotope[214]{Bi} and \isotope[208]{Tl} (daughter nuclei of \isotope[238]{U} and \isotope[232]{Th}, respectively). These results will both inform preparation for scintillator fill and support water phase physics analyses like the search for invisible nucleon decay modes.
\end{abstract}

\section{Introduction to SNO+}
SNO+ is a neutrino experiment with the main goal of searching for neutrinoless double-beta decay. It is located 2km underground at SNOLAB, in Sudbury, Canada. and consists of a 12m diameter acrylic sphere surround by around 9400 photomultiplier tubes (PMT) held by a PMT support structure. The whole setup is located in a 40m tall cylindrical cavity which is filled with ultrapure water (UPW) for additional shielding. SNO+ will be run in three phases: water phase, liquid scintillator phase and $0\nu\beta\beta$ decay search phase using \isotope[130]{Te}-loaded liquid scintillator \cite{andriga}. SNO+ is currently in the water phase, where the acrylic vessel and the surrounding cavity is filled with ultra-pure water. Data taking for the water phase has been ongoing since May 2017. \\

There are a few goals SNO+ aims to achieve in the water phase. One of these goals is to ensure that the hardware is in optimum condition for the scintillator phase. This will give SNO+ the opportunity to perform any necessary repairs and upgrades to the hardware before liquid scintillator is placed in the detector. In addition, the water phase enables initial calibration of the detector, such as optical calibrations. This gives us the opportunity to prepare and test the calibration hardware as well as making otherwise difficult measurements such as water and AV optical properties. It also enables us to measure backgrounds (eg: PMTs, acrylic vessel(AV) ) in the external water (region between the outer AV and PMTs). Finally, SNO+ aims to perform physics measurements such as a search for invisible nucleon decay modes. \\ 

\section{Nucleon Decay Search}
\begin{figure}[h]
\includegraphics[width=28pc]{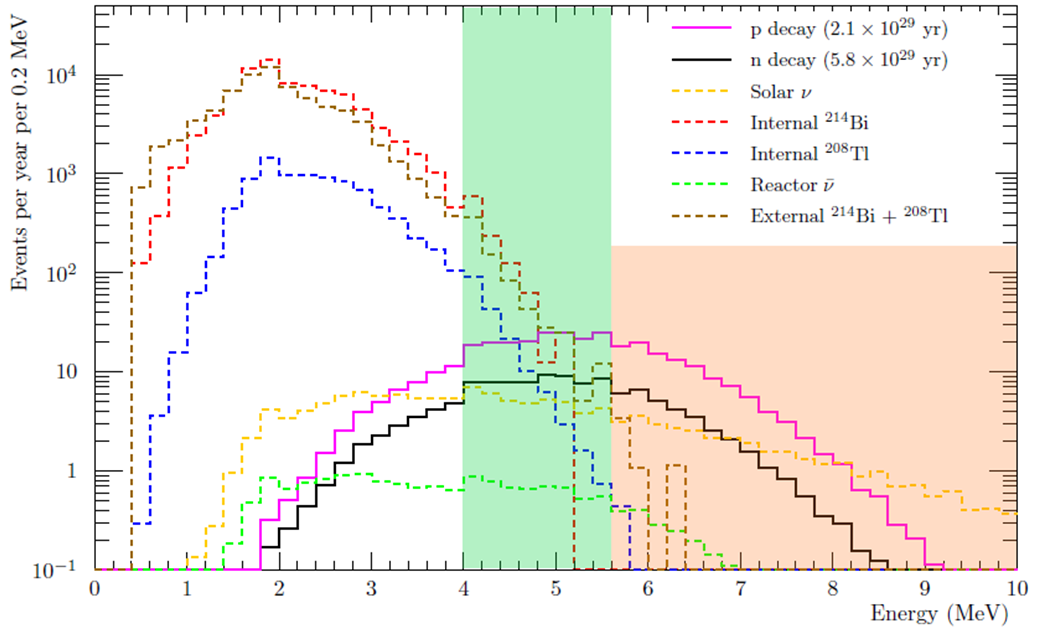}\hspace{2pc}%
\begin{minipage}[b]{0.2\textwidth}
\caption{\label{ND} The region of interest for the ND search(orange) is currently in blinded data. A sideband region (green) is used for the internal backgrounds analysis. The sideband region was chosen to avoid low energy instrumentals and low energy decays.}
\end{minipage}
\end{figure}

As stated above, the nucleon decay search is one of the major physics measurements of the water phase. In particular, we are looking at the ``invisible" nucleon decay channels. An example decay would be: \isotope[16]{O} $\rightarrow$ \isotope[15]{O}$^{*}$ + $\bar{\nu}\bar{\nu}\nu$ . The de-excitation of \isotope[15]{O} would create a gamma around 6-7 MeV around 45\% of the time. The current best limit is $\tau_{n} >$ 5.8$\times10^{29}$ years \cite{araki}. A simulation of a possible nucleon decay signal and expected backgrounds in SNO+ water phase is shown in \hyperref[ND]{Figure \ref*{ND}}. 

\section{Internal Backgrounds} 
\begin{figure}[h]
\begin{minipage}{0.5\textwidth}
\includegraphics[width=\textwidth]{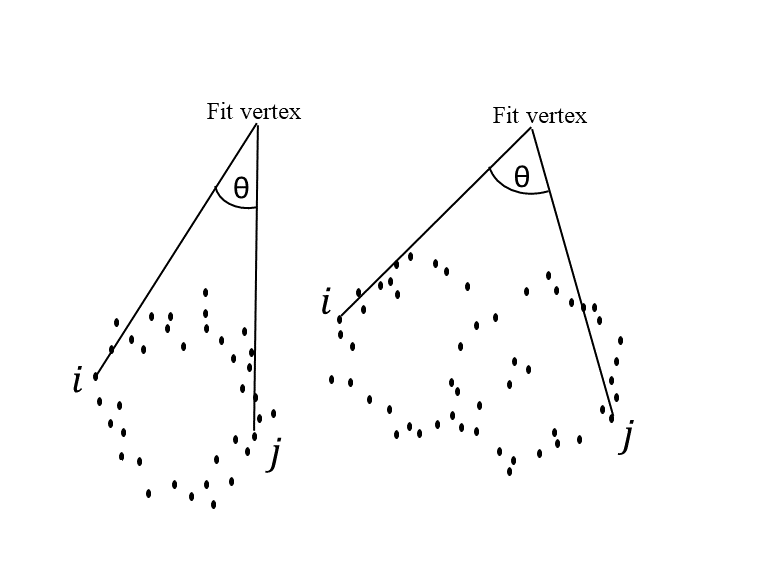}
\caption{\label{beta14_diag}Parameterization of the isotropy of detected light: $ \beta_{14} = \beta_{1} + 4\beta_{4}$, where $\beta_{l} = \frac{2}{N(N-1)}\left[\sum_{i=1}^{N-1}\sum_{j=i+1}^{N}P_{l}(cos\theta_{ij})\right]$ \cite{okeeffe}.}
\end{minipage}\hspace{2pc}%
\begin{minipage}{0.5\textwidth}
\includegraphics[width=\textwidth]{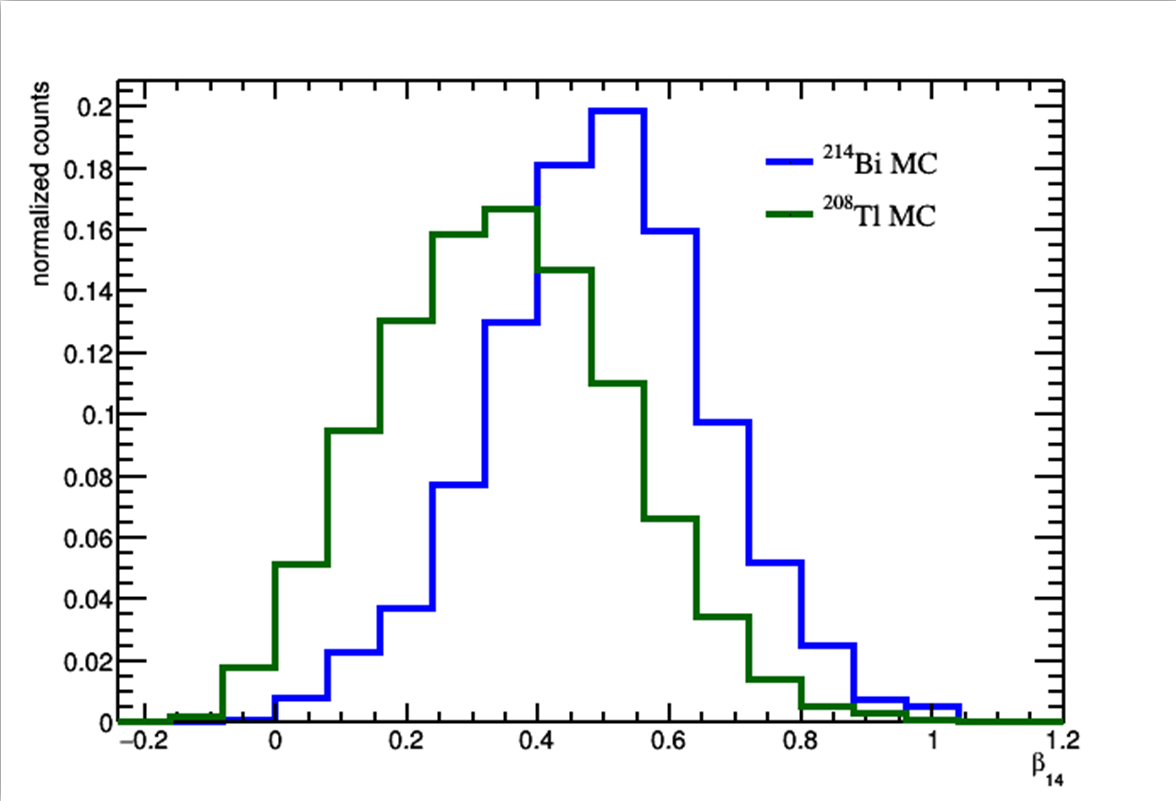}
\caption{\label{beta14_mc}A simulation showing the isotropy distribution for \isotope[214]{Bi} and \isotope[208]{Tl}. There is a clear difference in the $\beta_{14}$ spectrum between the two isotopes, with \isotope[208]{Tl} being more isotropic than \isotope[214]{Bi}. This was due to Tl emitting a 2.614 MeV $\gamma$ and a 1.8MeV end-point $\beta$ compared to Bi whose decays are dominated by a 3.27 MeV $\beta$.}
\end{minipage} 
\end{figure}

Since nucleon decay is a rare event, a good understanding of the internal backgrounds (backgrounds originating within the active/fiducial volume of the detector) is required. The two most important internal backgrounds, \isotope[214]{Bi} and \isotope[208]{Tl} are daughters of \isotope[238]{U} and \isotope[232]{Th}, respectively. We can discriminate them using the difference in isotropy of the Cherenkov light of the two isotopes, resulting from the fact that they have very different gamma cascades associated with their beta decays, as first applied in SNO \cite{dunmore}. The parameter $\beta_{14}$ is the isotropy parameter and it is a measure of the spread of hit PMTs with respect to the event fit vertex, shown in \hyperref[beta14_diag]{Figure \ref*{beta14_diag}}. The difference between \isotope[214]{Bi} and \isotope[208]{Tl} events in this parameter is illustrated in \hyperref[beta14_mc]{Figure \ref*{beta14_mc}}.\\

\section{Preliminary Results}
An energy sideband analysis carried out on a subset of the water data gives the following results, which are converted to equivalent parent concentration assuming secular equilibrium: \\

\begin{center}
\begin{tabular}{l}
gU/gH2O = $(5.4 \pm 0.7_{(stat)} \pm 2.9_{(sys)})\times 10^{-14}$ \\
gTh/gH2O = $< 2.9 \times 10^{-14}$ (95\% upper C.L.)
\end{tabular}
\end{center}

\begin{figure}[h]
\includegraphics[width=22pc]{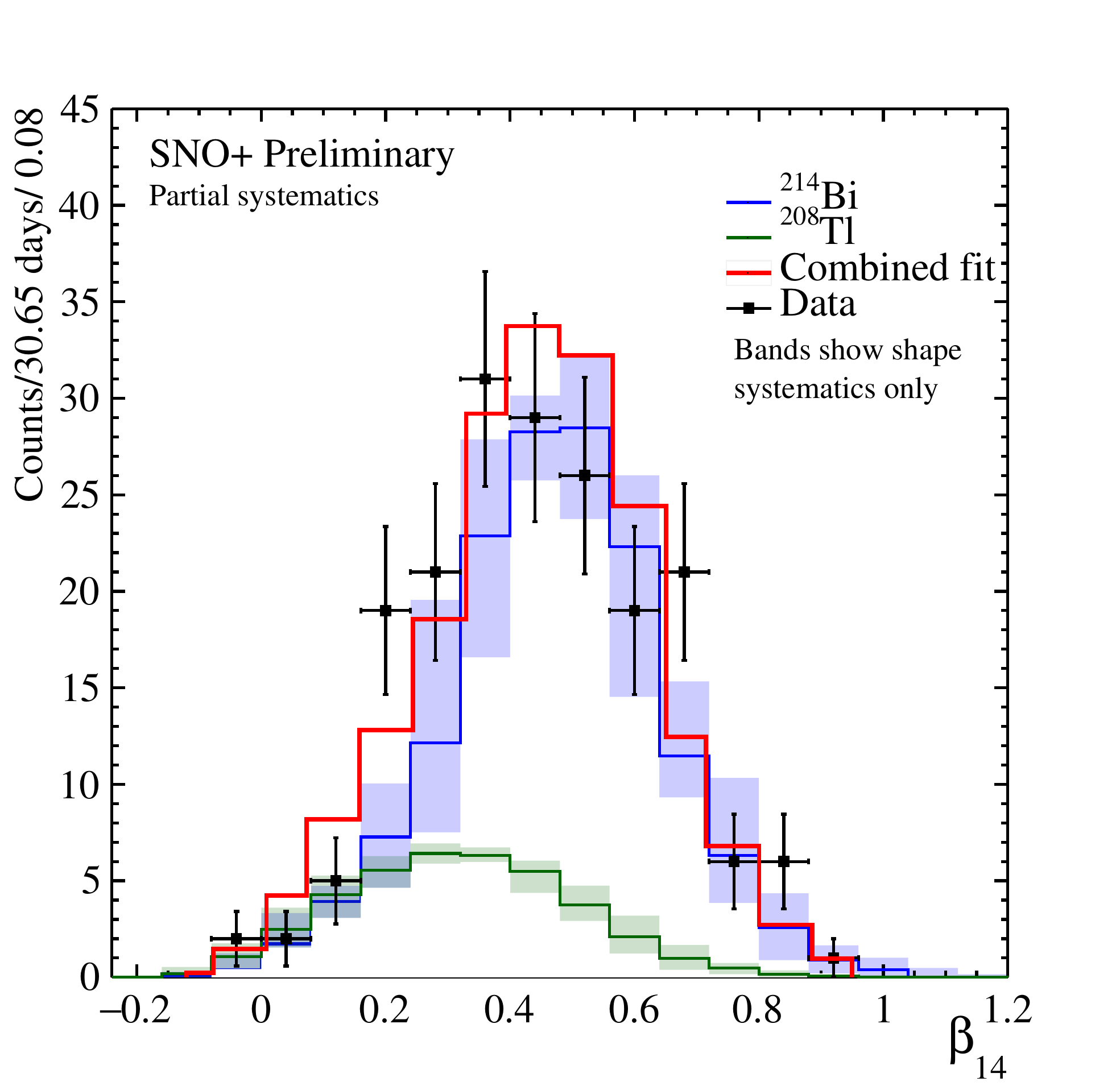}\hspace{0.5pc}%
\begin{minipage}[b]{0.4\textwidth}
\caption{\label{tb3}The fitted isotropy distribution of the events in the energy sideband for a subset of the water data. Fit results are shown above, and are expressed in U/Th equivalent concentration assuming secular equilibrium.}
\end{minipage}
\end{figure}

\section{Conclusions}
This proceeding shows that first measurements of the levels of internal backgrounds in the SNO+ water phase have been obtained. These results are consistent with our target levels for the nucleon decay analysis. \\

\ack
This work is supported by ASRIP, CIFAR, CFI, DF, DOE, ERC, FCT, FedNor, NSERC, NSF, Ontario MRI, Queens University, STFC, UC Berkeley and benefitted from services provided by EGI, GridPP and Compute Canada. The presenter is supported by NSERC and Ontario Early Researcher Awards Program. We thank SNOLAB and Vale for valuable support.

\end{document}